\documentclass[aps,prb,reprint,groupedaddress]{revtex4-1}
\usepackage{dcolumn}% Align table columns on decimal point
\usepackage{bm}% bold math
\usepackage[dvipdfmx]{graphicx}
\usepackage{color}
\usepackage{array}
\usepackage{mathrsfs}
\usepackage{ulem} 
\usepackage{amsmath}
\usepackage{amssymb}
\usepackage{url}

\begin{document}

% Use the \preprint command to place your local institutional report
% number in the upper righthand corner of the title page in preprint mode.
% Multiple \preprint commands are allowed.
% Use the 'preprintnumbers' class option to override journal defaults
% to display numbers if necessary
%\preprint{}

%Title of paper
\title{Finite-temperature properties of the Kitaev-Heisenberg models on kagome and triangular lattices studied by improved finite-temperature Lanczos methods}

\author{Katsuhiro Morita}
\email[e-mail:]{katsuhiro.morita@rs.tus.ac.jp}
\affiliation{Department of Applied Physics, Tokyo University of Science, Tokyo 125-8585, Japan}

\author{Takami Tohyama}
\affiliation{Department of Applied Physics, Tokyo University of Science, Tokyo 125-8585, Japan}

\date{\today}

\begin{abstract}
Frustrated quantum spin systems such as the Heisenberg and Kitaev models on various lattices, have been known to exhibit various exotic properties not only at zero temperature but also for finite temperatures.
Inspired by the remarkable development of the quantum frustrated spin systems in recent years,
we investigate the finite-temperature properties of the $S=1/2$ Kitaev-Heisenberg models on kagome and triangular lattices by means of finite-temperature Lanczos methods with improved accuracy.
In both lattices, multiple peaks are confirmed in the specific heat.
To find the origin of the multiple peaks, we calculate the static spin structure factor.
The origin of the high-temperature peak of the specific heat is attributed to a crossover from the paramagnetic state to a short-range ordered state whose static spin structure factor has zigzag or linear intensity distributions in momentum space.
In the triangular Kitaev model, the ``order by disorder'' due to quantum fluctuation occurs. On the other hand, in the kagome Kitaev model it does not occur even with both quantum and thermal fluctuations. 
\end{abstract}
\pacs{}

\maketitle

\section{INTRODUCTION}
\label{Sec1}
The $S=1/2$ antiferromagnetic Heisenberg models on triangular lattice (TL) and kagome lattice (KL), which have strong geometric frustration arising from triangle units with antiferromagnetic interaction, have been studied for over several decades both experimentally~\cite{te1,te2,te3,te4, ke1,ke2,ke3,ke4,ke5,ke6,ke7,ke8,ke9,ke10,ke11,ke12,ke13,ke15,ke16} and theoretically~\cite{t1201,t1202,t1203,tft1,tft2,tft3,kt1,kt2,kt3,kt4,kg1,kg2,kg3,
kg4,kg5,kg6,ku1,ku2,ku3,ku4,ku5,kv1,kv2,kv3,kv4,kft1,kft2,kft3,kft4,kft5,kft6,kft7,kft8}.
The strong frustration prevents collinear-type magnetic orders in their ground states.
In the TL, the ground state exhibits $120^\circ$ order~\cite{t1201,t1202,t1203}, whereas in the KL it is predicted to be the quantum spin liquids~\cite{kg1,kg2,kg3,kg4,kg5,kg6,ku1,ku2,ku3,ku4,ku5} 
or valence bond crystals~\cite{kv1,kv2,kv3,kv4}.
At finite temperature, these models commonly show multiple-peak structures in the temperature dependence of specific heat owing to the frustration effect~\cite{tft3,kft1,kft2,kft3,kft4,kft6,kft8}.

The $S=1/2$ Kitaev model on the honeycomb lattice (HL) dose not have geometric frustration but has frustration effects arising from the bond-dependent Ising interactions~\cite{hk}, called exchange frustration.
In this model, the $S=1/2$ spins are divided into localized Majorana fermions composing $Z_2$ fluxes and itinerant Majorana fermions~\cite{hkm1,hkm2,hkm3}.
Its ground state exhibits an exact quantum spin liquid with topological order. At finite temperatures, there is a distinct double peak in the specific heat~\cite{hkdp}.
The origin of this double peak is described below: the high-temperature peak is caused by freezing the itinerant Majorana fermions and the low-temperature peak is caused by freezing the localized Majorana fermions~\cite{hkdp}.
Because of clear difference of their energy scales, a 1/2-plateau-like anomaly appears in the temperature dependence of the entropy. 
This phenomenon corresponds to a fractional excitation of the spins.
Moreover, such a phenomenon has been found even in the spin $S>1/2$ and mixed spin systems~\cite{hkas1,hkas2}, even though the spin degree of freedom cannot be decomposed into Majorana fermions. 
Furthermore, finite-temperature properties of the Kitaev-Heisenberg (KH) model have also been studied on the HL~\cite{hkh1,hkh2}. 

The $S=1/2$ KH models on the KL and TL, having both the geometric frustration and exchange frustration, have been studied mainly for the ground state~\cite{kkh,tkh1,tkh2,tkh3,tkh4,tkh5,tkh6}.
In the KL-KH system, it has been proposed that there are two quantum spin liquids, a canted ferromagnetic, and the ${\bf q=0 }$, 120$^\circ$ ordered phases~\cite{kkh},
whereas in the TL-KH system, it has been proposed that there are $Z_2$ vortex crystal, nematic, dual-$Z_2$ vortex crystal, ferromagnetic, and dual-ferromagnetic phases~\cite{tkh2,tkh3,tkh4,tkh5,tkh6}.
However, finite-temperature properties in the KH models on the KL and TL have hardly been investigated.
There is a possibility that multiple peaks in the temperature dependence of the specific heat and new crossover phenomena exist, because such phenomena have been confirmed in the HL-Kitaev and KL-Heisenberg models.
Therefore, it is important to investigate the finite-temperature properties of these models.

The finite-temperature Lanczos method (FTLM) is a useful technique for calculating finite-temperature properties~\cite{ftl1,ftl2}.
However, this method has a problem that the accuracy becomes worse at low temperatures~\cite{ftl2}. Therefore, we need to overcome this problem.
In this paper, we first propose two methods to improve the FTLM.
We name the methods the replaced finite-temperature Lanczos method (RFTLM) and orthogonalized finite-temperature Lanczos method (OFTLM).
Using these improved FTLMs, we next calculate the specific heat,  entropy, and static spin structure factor (SSSF) to investigate the finite-temperature properties of the $S=1/2$ KH model on the KL and TL.
In the kagome system, the specific heat exhibits multiple-peak structures at finite temperatures for $0  \leq \theta \leq 0.5\pi$, where $\theta=\arctan(K/J)$ with $J$ ($K$) being the Heisenberg (Kitaev) interaction.
To clarify the origin of the multiple-peak structure of the specific heat, we analyze the SSSF at finite temperatures for the $N=36$ cluster using the RFTLM. 
From the analyses, we find that the highest-temperature peak of the specific heat for $0.1\pi  \leq \theta \leq 0.4\pi$ originates with a crossover from the paramagnetic state to a state whose SSSF intensity shows direction distribution in the momentum space.
On the other hand, one of the low-temperature peaks for $0.1\pi  \leq \theta \leq 0.4\pi$ is expected to be a signature of the emergence of a $\bf q=0$, $120^\circ$ order state.
However, at $\theta=0.5\pi$ (Kitaev limit), the $\bf q=0$, $120^\circ$ order does not appear.
In the triangular system, we find that there is a double-peak structure in the specific heat for $0.25\pi \leq \theta \leq 0.5\pi$.
The origin of the double-peak structure is the same as the kagome system.
At $\theta=0.5\pi$, the ground state exhibits a stripe order due to the ``order-by-disorder'' mechanism unlike the kagome system.

The arrangement of this paper is as follows. In Sec.~\ref{Sec2}, we describe our $S=1/2$ KH models on KL and TL. In Sec.~\ref{Sec3},  we first explain the standard FTLM; then we explain the RFTLM and OFTLM developed by us. 
In Sec.~\ref{Sec4}, the results of the specific heat, entropy, and SSSF for the KL and TL are shown. 
In Sec.~\ref{Sec5}, we discuss the difference between the honeycomb, kagome, and triangular systems for the origin of the multiple-peak structures in the specific heat and we focus on characteristic of the Kitaev model on the KL.
Finally, a summary is given in Sec.~\ref{Sec6}.

\begin{figure}[tb]
 \begin{center}
	\includegraphics[width=80mm]{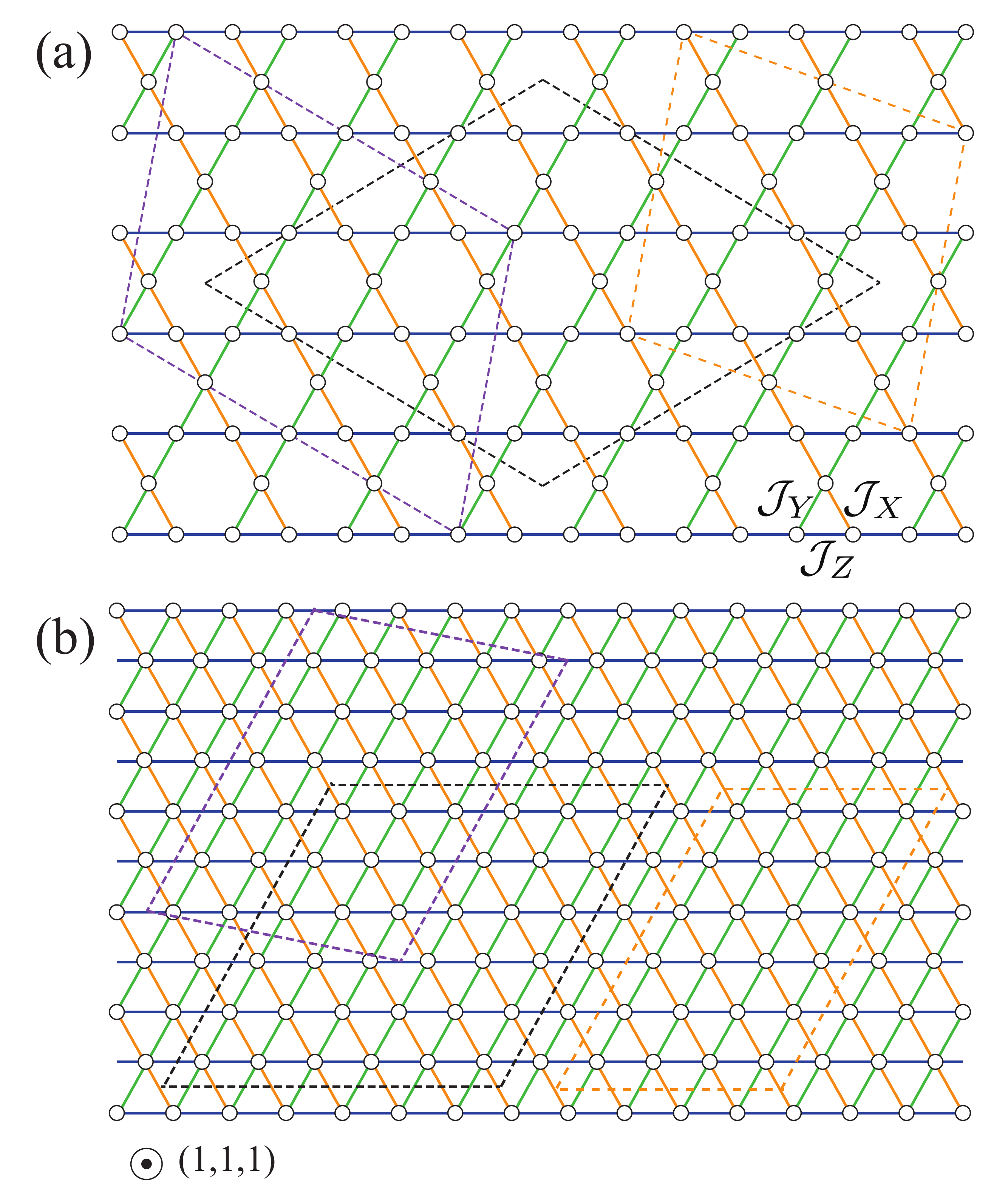}
   \caption{Lattice structure of the KL (a)  and TL (b) with three anisotropic exchange interactions, $\mathcal{J}_X$, $\mathcal{J}_Y$, and $\mathcal{J}_Z$. 
The orange, green, and blue solid lines denote $\mathcal{J}_X$, $\mathcal{J}_Y$, and $\mathcal{J}_Z$, respectively. 
The orange, purple, and black dashed quadrangles denote the clusters of $N=24$, $N=30$, and $N=36$, respectively, used in the FTLMs with periodic boundary conditions.}
  \label{lattice}
\end{center}
\end{figure}

\section{model}
\label{Sec2}

The Hamiltonian of the KH model is given by
\begin{equation}
  \mathcal{H}= \sum _{\langle i,j \rangle} {\bf S}_i ^{\rm T}  \mathcal{J}_{i,j}{\bf S}_j \label{hamiltonian},
\end{equation}
where ${\bf S}_i$ is a quantum spin operator with $S=1/2$ at site $i$. $\mathcal{J}_{i,j}$ represents the nearest neighbor interactions as shown in Fig.~\ref{lattice}(a) for the KL and 
 Fig.~\ref{lattice}(b) for the TL. $\mathcal{J}_{i,j}$ takes one of the three anisotropic interactions, $\mathcal{J}_X=\mathrm{diag}(J+K, J, J)$ (yellow bonds), $\mathcal{J}_Y = \mathrm{diag}(J, J+K, J)$ (light green bonds), and $\mathcal{J}_Z=\mathrm{diag}(J, J, J+K)$ (blue bonds), where $K$ and $J$ correspond to the energy of the Kitaev and Heisenberg interactions, respectively.  
We introduce the parametrization $(J,K) = (I\cos\theta,I\sin\theta)$, where $I$ is the energy unit ($I=1$).
In the present study, we focus on $0 \leq \theta \leq 0.5\pi$.

\section{methods}
\label{Sec3}
\subsection{Finite-temperature Lanczos method}
In this section, we describe the standard FTLM~\cite{ftl1,ftl2}.
The FTLM has been used to study the finite-temperature properties of various lattice models~\cite{kft3,kft6,ftla1,ftla2,ftla3,ftla4,ftla5,ftla6,ftla7,ftla8,ftla9,ftla10}.
The partition function $Z(T)$ of the canonical ensemble at temperature $T$ is expressed as follows:
\begin{eqnarray} 
Z(T) &=& \sum _{n=1}^{N_{st}} \langle n | e^{-\beta  \mathcal{H}} | n\rangle \nonumber \\
	    &=& \sum _{n=1}^{N_{st}}  \sum _{i=0}^{N_d-1} \sum _{k=1}^{d_i}e^{-\beta E_i} \langle n | \Psi_{ik} \rangle \langle \Psi_{ik} | n\rangle  \label{Z} \\
		&=& \sum _{i=0}^{N_d-1} d_i e^{-\beta E_i}  \label{Z2},
\end{eqnarray}
where $N_{st}$ is the dimension of $\mathcal{H}$, $|n\rangle$ is an arbitrary normalized vector, $\beta$ is the inverse temperature 1/$T$ ($k_B=1$), $E_i$ is an eigenenergy of $\mathcal{H}$,  $| \Psi_{ik} \rangle$ is an eigenvector with $E_i$, $d_i$ is a degree of degeneracy of the state with $E_i$, and $N_d$ represents the number of the eigenenergies, which satisfies $N_{st}=\sum _{i=0}^{N_d-1} d_i$.
The FTLM introduces two approximations for (\ref{Z}).
The first one is to replace the summation of $n$ with random sampling $r$ with $R$ times.
The second one is for the summations of $i$ and $k$. 
Both the summations are replaced by the Krylov subspace with dimension $M$. 
In the FTLM, the partition function and general operator $A$ are approximated as follows:
\begin{eqnarray} 
 Z(T)_{\rm FTL} &=& \frac{N_{st}}{R}\sum _{r=1}^{R} \sum _{j=0}^{M-1}  e^{-\beta \epsilon^{(r)}_j} |\langle V_r | \psi^r_j \rangle|^2,  \label{ZFTL} \\
 \langle A \rangle(T)_{\rm FTL}\! &=& \!\frac{N_{st}}{RZ(T)_{\rm FTL}}\!\sum _{r=1}^{R}\!\sum _{j=0}^{M-1}\!e^{-\beta \epsilon^{(r)}_j}\!\langle V_r | \psi^r_j \rangle\!\langle \psi^r_j | A | V_r \rangle,   \label{A}
\end{eqnarray}
where $|V_r\rangle$ is a normalized random initial vector 
and $| \psi^r_j \rangle$ ($\epsilon^{(r)}_j$) are an eigenvector (eigenvalue) in the $M$-th Krylov subspace for $\mathcal{H}$. 
We note that $|V_r\rangle$ is formally given by $|V_r\rangle = \sum_{i=0}^{N_d-1}\sum_{k=1}^{d_i} \eta_{rik}|\Psi_{ik} \rangle$ using the exact eigenstate $|\Psi_{ik} \rangle$, where $\eta_{rik}$ is a random value that satisfies $\sum_{i=0}^{N_d-1}\sum_{k=1}^{d_i} |\eta_{rik}|^2=1$ for the normalization.

For the energy $E(T)$, specific heat $C(T)$, and entropy $\mathcal{S}(T)$, the following general expressions are useful: $E(T) = -\frac{\partial}{\partial \beta} \ln Z(T)$, $C(T) = \frac{\partial}{\partial T} E(T)$, and  $\mathcal{S}(T) = \frac{E(T)}{T} + \ln Z(T)$.
From these equations, $E(T)$ and $C(T)$ calculated by the FTLM are given by
\begin{eqnarray} 
	E(T)_{\rm FTL} &=&  \frac{N_{st}}{RZ(T)_{\rm FTL}}\sum _{r=1}^{R} \sum _{j=0}^{M-1}  \epsilon^{(r)}_j e^{-\beta \epsilon^{(r)}_j} |\langle V_r | \psi^r_j \rangle|^2, \\
    C(T)_{\rm FTL}  &=&  \frac{N_{st}}{T^2RZ(T)_{\rm FTL}}\sum _{r=1}^{R} \sum _{j=0}^{M-1} |\epsilon^{(r)}_j|^2 e^{-\beta \epsilon^{(r)}_j} |\langle V_r | \psi^r_j \rangle|^2   \nonumber \\
           & &       -\frac{|E(T)_{\rm FTL}|^2}{T^2}. \label{C}
\end{eqnarray}
At high temperatures, $R$ of a few samplings is enough for obtaining high accuracy since the error of all physical quantities is proportional to $\mathcal{O}(1/\sqrt{RN_{st}})$~\cite{ftl2} with a large number of $N_{st}$. 
On the other hand, for $T\rightarrow0$, $C(T\rightarrow0)_{\rm FTL}$ and $E(T\rightarrow0)_{\rm FTL}$ reach an exact value if $|\psi^r_0 \rangle$ becomes a ground state $|\Psi_0^r \rangle$~\cite{ftl2},
where $|\Psi_0^r \rangle = \sum_{k=1}^{d_0} \eta_{r0k}|\Psi_{0k} \rangle/\sqrt{\sum_{k=1}^{d_0} |\eta_{r0k}|^2}$.
 $\mathcal{S}(T\rightarrow0)_{\rm FTL}$ and $\langle A \rangle (T\rightarrow0)_{\rm FTL}$ read
\begin{eqnarray} 
  \mathcal{S}(T\rightarrow0)_{\rm FTL} &=& \ln\frac{N_{st}}{R}\sum _{r=1}^{R} |\langle V_r | \Psi_0^r \rangle|^2, \label{S0} \\
\langle A\rangle(T\rightarrow0)_{\rm FTL}&=& \frac{\sum _{r=1}^{R} \langle V_r | \Psi_0^r \rangle  \langle \Psi_0^r | A | V_r \rangle}{\sum _{r=1}^{R} |\langle V_r | \Psi_0^r \rangle|^2 }.  \label{A0} 
\end{eqnarray}
Equation (\ref{S0}) does not give an exact value, and
if $A$ is noncommutative with Hamiltonian such as the SSSF, Eq.~(\ref{A0}) also does not give an exact value.
These errors are expected to be $\mathcal{O}(1/\sqrt{R})$~\cite{ftl2}.
Therefore, a very large number of samplings is required to obtain good accuracy at low temperatures.
The low-temperature Lanczos method~\cite{lftl} is known as one of the solutions to this problem. However, this method has a difficulty for large-scale calculations because it requires huge random access memory to keep all vectors in the Krylov subspace with $M$.
Therefore, we try to improve the accuracy of the FTLM at low temperature in two ways: the RFTLM and OFTLM.

\subsection{Replaced finite-temperature Lanczos method (RFTLM)}
In the standard Lanczos method, we can obtain several low-lying eigenstates with $N_E$ levels whose energy is given by $\epsilon^{(r)}_i$ ($i=0, 1, \cdots, N_E-1$), but we cannot judge the degeneracy of each level. Therefore, $\epsilon^{(r)}_0 < \epsilon^{(r)}_1 < \cdots < \epsilon^{(r)}_{N_E-1}$ and each eigenvector should be written generally $|\Psi_i^r \rangle = \sum_{k=1}^{d_i} \eta_{rik}|\Psi_{ik} \rangle/\sqrt{\sum_{k=1}^{d_i} |\eta_{rik}|^2}$ using $d_i$-fold-degenerate exact eigenvector $|\Psi_{ik} \rangle$.
Here, we assume that the obtained energy $\epsilon^{(r)}_i$ is independent of sampling $r$, i.e., $\epsilon^{(r)}_i=E_i$, although the corresponding eigenvector may depend on the sampling
$|\psi^r_i \rangle= | \Psi^r_i \rangle$ due to possible degeneracy. 
Then we can rewrite expression (\ref{ZFTL}) as follows:
\begin{eqnarray} 
  Z(T)_{\rm FTL}    &=&  \frac{N_{st}}{R}\sum _{r=1}^{R} \sum _{i=0}^{N_E-1}  e^{-\beta E_i} |\langle V_r | \Psi^r_i \rangle|^2 \nonumber  \\
            &+&  \frac{N_{st}}{R}\sum _{r=1}^{R} \sum _{j=N_E}^{M-1}  e^{-\beta \epsilon^{(r)}_j} |\langle V_r | \psi^r_j \rangle|^2. \label{ZRFTL} 
\end{eqnarray}
Comparing the first term on the right-hand side of Eq.~(\ref{ZRFTL}) with Eq.~(\ref{Z2}), we come up with replacing

\begin{eqnarray} 
\langle V_r | \Psi^r_i \rangle \Rightarrow \sqrt{\frac{d_i}{N_{st}}}. \label{rep}
\end{eqnarray}
The replacement (\ref{rep}) leads to the partition function of the RFTLM
\begin{eqnarray} 
  Z(T)_{\rm RFTL}    &=& \sum _{i=0}^{N_E-1} d_i e^{-\beta E_i}  \nonumber  \\
            &+&  \frac{N_{st}}{R}\sum _{r=1}^{R} \sum _{j=N_E}^{M-1}  e^{-\beta \epsilon^{(r)}_j} |\langle V_r | \psi^r_j \rangle|^2. \label{ZRFTL2} 
\end{eqnarray}
The first term in Eq.~(\ref{ZRFTL2}) is the same as the exact partition function $Z(T)$ (\ref{Z2}), for $i <N_E$.
This indicates that $Z(T)_{\rm RFTL}$ (\ref{ZRFTL2}) is more accurate than $Z(T)_{\rm FTL}$ (\ref{ZFTL}).
In a similar way, $\langle A \rangle(T)_{\rm FTL}$ can be improved in accuracy by replacing  
\begin{eqnarray} 
\langle \Psi^r_i | A | V_r \rangle \Rightarrow \frac{1}{\sqrt{d_iN_{st}}}\sum _{k=1}^{d_i} \langle \Psi_{ik} | A | \Psi_{ik} \rangle
\end{eqnarray}
for $i<N_E$. 
$\langle A \rangle(T)$ using RFTLM reads
\begin{eqnarray} 
 \langle A \rangle(T)_{\rm RFTL} = \frac{1}{Z(T)_{\rm RFTL}}    \sum_{i=0}^{N_E-1} e^{-\beta E_i}  \sum_{k=1}^{d_i}  \langle \Psi_{ik} | A | \Psi_{ik} \rangle \nonumber \\
+ \frac{N_{st}}{RZ(T)_{\rm RFTL}}\sum _{r=1}^{R} \sum _{j=N_E}^{M-1}  e^{-\beta \epsilon^{(r)}_j} \langle V_r | \psi^{r}_j \rangle  \langle \psi^{r}_j | A | V_r \rangle.  \label{ARFTL} 
\end{eqnarray}

We can obtain the exact eigenstates $|\Psi_{ik}\rangle$ with $E_i$ by the several kinds of exact diagonalization (ED) methods such as the thick-restart Lanczos method~\cite{trl}, band Lanczos method~\cite{blm}, locally optimal block preconditioned conjugate gradient method~\cite{lob}, and root-shifting method~\cite{roo}.

By performing the RFTLM, $\mathcal{S}(T\rightarrow0)_{\rm RFTL}$ and $\langle A \rangle(T\rightarrow0)_{\rm RFTL}$ become an exact value $\ln(d_0)$ and $\sum _{k=1}^{d_0} \langle \Psi_{0k} | A | \Psi_{0k}\rangle/d_0$, respectively.
Therefore, accuracy at low temperatures using the RFTLM would be extremely improved as compared with the standard FTLM.
The efficacy of the RFTLM is confirmed in Sec.~\ref{ben}.

However, in the RFTLM, it is necessary to know the degeneracy $d_i$ in order to perform the summation of $i$ in Eqs.~(\ref{ZRFTL2}) and (\ref{ARFTL}).
We also should be careful about pseudo-eigenvalues, so-called ``ghost'' eigenvalues caused by the presence of the machine epsilon.
If there are ghost eigenvalues, it is necessary to change $N_E$ in the second term of Eqs.~(\ref{ZRFTL2}) and (\ref{ARFTL}) to $N_E + N_g$, where $N_g$ is the number of the ghost eigenvalues less than $E_{N_E}$.
We develop a new method in the next section to overcome these problems.

\subsection{Orthogonalized finite-temperature Lanczos method (OFTLM)}
In this subsection, for simplicity, we include the index $k$ for degeneracy into the index $i$ hereafter, rewriting  $\eta_{rik} \Rightarrow \eta_{ri}$ and $|\Psi_{ik}\rangle \Rightarrow |\Psi_{i}\rangle$.
Thus the random vector $|V_{r}\rangle$ reads $|V_r\rangle = \sum_{i=0}^{N_{st}-1} \eta_{ri}|\Psi_{i} \rangle$.
In the OFTLM, we first calculate several low-lying exact eigenvectors $|\Psi_i\rangle$ with $N_V$ levels ($E_0 \leq E_1 \leq \cdots \leq E_{N_V-1}$) before performing the FTLM.
We next use the following modulated random vector:
\begin{eqnarray} 
 |V_r'\rangle &=& \sum_{i=N_V}^{N_{st}-1} \eta_{ri}|\Psi_{i} \rangle \nonumber \\ 
               &=& \left[ I - \sum_{i=0}^{N_V-1} | \Psi_{i} \rangle \langle \Psi_{i} |  \right] | V_r \rangle  \label{r'}
\end{eqnarray}
with normalization
\begin{equation}
 |V_r'\rangle  \Rightarrow \frac{ |V_r'\rangle }{ \sqrt{\langle V_r' |V_r'\rangle} }. \label{r'2}
\end{equation}
Here, $|V_r'\rangle$ is orthogonal to the states $|\Psi_{i} \rangle$ for $i < N_V$.
Therefore, the FTLM using $|V_r'\rangle$ as the initial vector is equivalent to applying the method to a Hilbert space excluding $|\Psi_{i}\rangle$ through $\sum_{i=0}^{N_V-1} | \Psi_{i} \rangle \langle \Psi_{i} | $, which has $N_{st}-N_V$ dimensions.
$Z(T)$ and  $\langle A \rangle(T)$ of the OFTLM are obtained by adding exact values coming from $|\Psi_i\rangle$ to the FTLM result obtained by using $|V_r'\rangle$ as an initial vector:
\begin{eqnarray} 
  Z(T)_{\rm OFTL}  &=& \frac{N_{st}-N_V}{R}\sum _{r=1}^{R} \sum _{j=0}^{M-1}  e^{-\beta \epsilon^{(r)}_j} |\langle V_r' | \psi^{r}_j \rangle|^2  \nonumber \\ 
          &+& \sum_{i=0}^{N_V-1}  e^{-\beta E_i}, \label{ZOFTL} 
\end{eqnarray}
\begin{eqnarray} 
 \langle A \rangle(T)_{\rm OFTL} &=& \frac{N_{st}-N_V}{RZ(T)_{\rm OFTL}}\sum _{r=1}^{R} \sum _{j=0}^{M-1}  e^{-\beta \epsilon^{(r)}_j} \langle V_r' | \psi^{r}_j \rangle  \langle \psi^{r}_j | A | V_r' \rangle \nonumber \\
                           &+&      \frac{1}{Z(T)_{\rm OFTL}}    \sum_{i=0}^{N_V-1}  e^{-\beta E_i}     \langle \Psi_{i} | A | \Psi_{i} \rangle.    \label{AOFTL}
\end{eqnarray}
Since $| \Psi_{i} \rangle$ obtained by the ED methods would be slightly different from the exact vectors because of the machine epsilon, 
some of the $\epsilon^{(r)}_j$ in the FTLM using $|V_r'\rangle$ may become, for example,  $E_0$, which should not appear.
In practical use, this is no problem since $|\langle V_r' | \psi^{r}_j \rangle|$ for such an $E_0$ becomes extremely small ($\sim$ machine epsilon).  
We can see that $Z(T)_{\rm OFTL}$ and $\langle A \rangle(T)_{\rm OFTL}$ are close to the exact values at low temperatures.
We emphasize that in the OFTLM we do not need to know the degeneracy $d_i$ in $|\Psi_{i}\rangle$ and can make $M$ smaller compared to the FTLM and RFTLM.
The efficacy of the OFTLM is confirmed in Sec.~\ref{ben}.

We note that an approach similar to the OFTLM has been discussed in terms of the kernel polynomial method.~\cite{sim}
\subsection{Confirming the efficacy of the RFTLM and OFTLM}
\label{ben}
\begin{figure*}[tb]
 \begin{center}
\includegraphics[width=154mm]{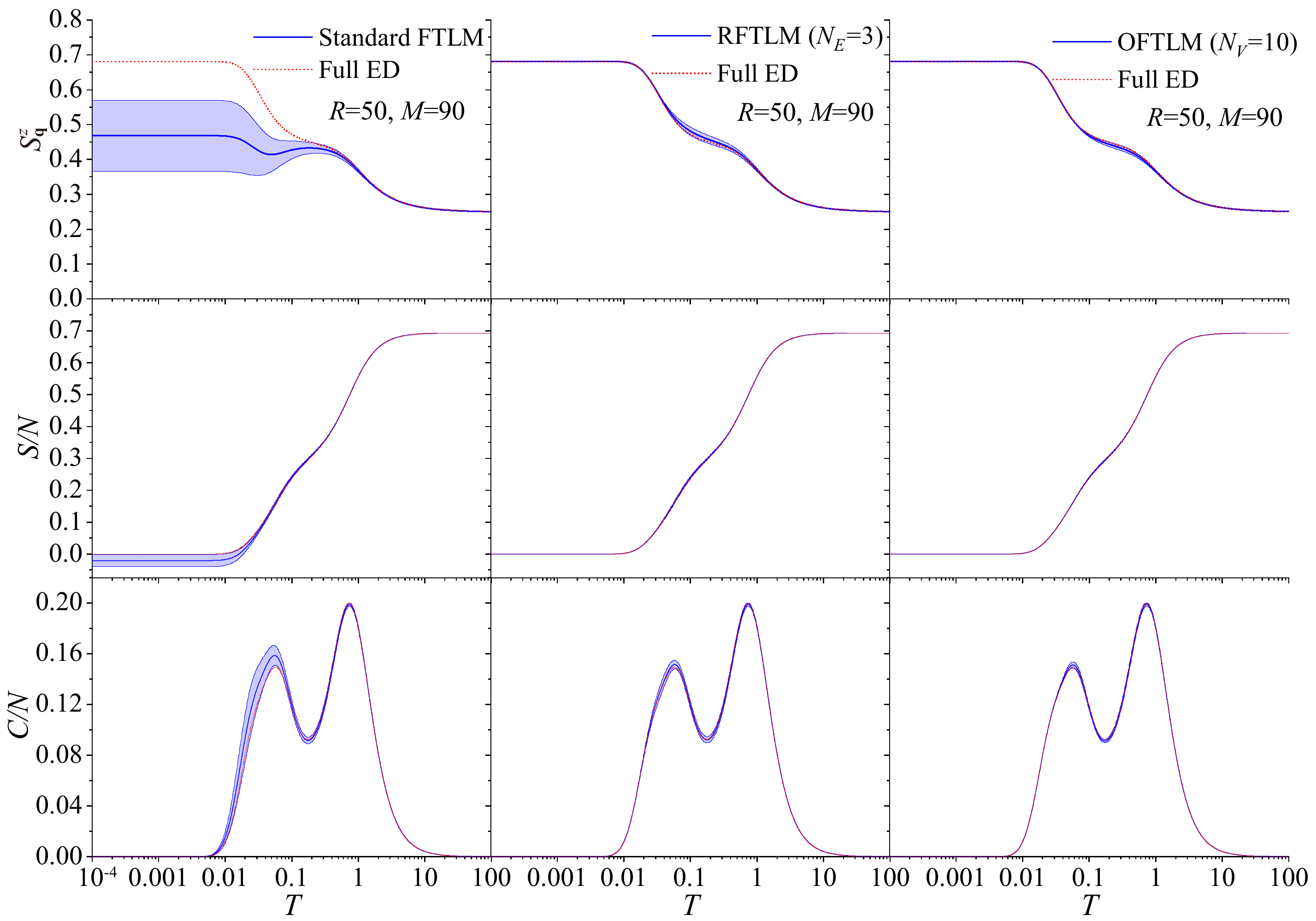}
   \caption{The accuracy of the FTLMs for the kagome system of $N=12$ at $\theta=0.2\pi$. 
The upper row, middle row, and lower row panels show $S^z_{\bf q}(T)$ at ${\bf q}=( 2\pi, 2\pi/\sqrt{3})$, $\mathcal{S}(T)/N$, and $C(T)/N$, respectively.
The left, middle, and right panels show the results using the standard FTLM, RFTLM with $N_E=3$, and OFTLM with $N_V=10$, respectively.
All the red dotted lines indicate the exact values using full ED.
The blue shaded regions indicate the standard errors of the FTLMs using the jackknife technique.
}
  \label{check1}
\end{center}
\end{figure*}

We perform benchmark calculations for the standard FTLM, RFTLM, and OFTLM.
We calculate $\mathcal{S}(T)$, $C(T)$, and the $z$ component of SSSF, $S^z_{\bf q}(T) = \langle S^z_{\bf q} \rangle(T)$, for an $N=12$ ($2\times2\times3$) kagome system with $\theta=0.2\pi$, where $S^z_{\bf q} =  \frac{1}{N}\sum_{j} \sum_{k} \it{e}^{\it{i}{\bf q}\cdot ({\bf r}_j-{ \bf r}_k)}S^z_{{\bf r}_j} S^z_{{\bf r}_k}$  with the position vector $\mathbf{r}_j$ and $\mathbf{r}_k$.
All FTLMs are performed with $M=90$ and $R=50$.
Here, we note that $M=90$ is large enough to obtain the ground state of the $N=12$ kagome system.
The calculated results are shown in Fig.~\ref{check1}.
The standard errors of the FTLMs using the jackknife technique~\cite{jk} are represented by the blue shaded regions in Fig.~\ref{check1}.
We can see that the accuracies of the RFTLM and OFTLM are clearly better than that of the standard FTLM for all physical quantities.
Therefore, we succeed in improving the FTLM.

\begin{figure*}[tb]
 \begin{center}
\includegraphics[width=156mm]{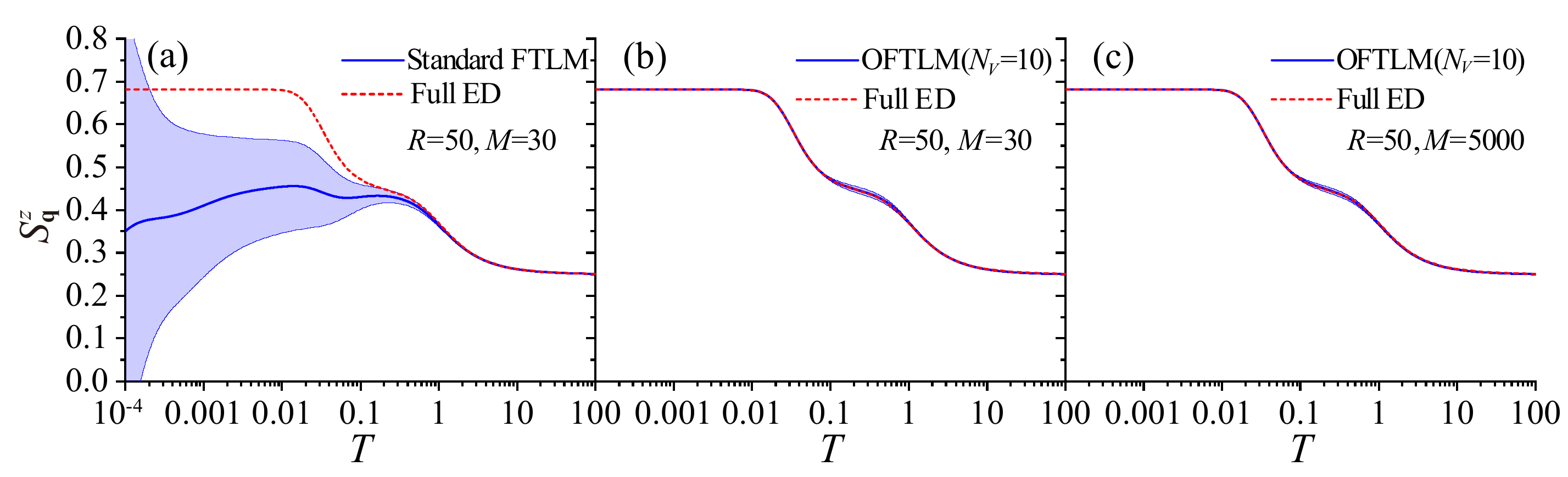}
   \caption{ The accuracy of $S^z_{\bf q}(T)$ at ${\bf q}=( 2\pi, 2\pi/\sqrt{3})$ using the FTLMs with respect to $M$ for the $N=12$ kagome system at $\theta=0.2\pi$.
All the red dashed lines indicate the exact values using full ED.
The blue shaded regions indicate the standard errors of the FTLMs using the jackknife technique.}
  \label{check2}
\end{center}
\end{figure*}

Furthermore, we compare the standard FTLM and OFTLM in detail using $S^z_{\bf q}(T)$ in Fig.~\ref{check2}.
In the standard FTLM, the accuracy for $M=30$ is very poor at low temperatures as shown in Fig.~\ref{check2}(a) because of small $M$ that is not enough to make a convergence to the ground state.
On the other hand, high-precision results can be achieved in the OFTLM even for the same $M$ [see Fig.~\ref{check2}(b)], 
since the contributions from low-energy sectors are added separately as shown in Eqs. (\ref{ZOFTL}) and (\ref{AOFTL}).
For this reason, the OFTLM gives a good convergence quicker then other FTLMs. 
In the OFTLM with larger $M$, the eigenvalues less than $E_{N_V}$ and the ghost eigenvalues appearing in the first terms of  Eqs. (\ref{ZOFTL}) and (\ref{AOFTL}) may affect $S^z_{\bf q}(T)$.
In order to investigate these effects, we also perform the OFTLM with very large $M=5000$ ($>N_{st}$).
We can see that there is no effect on $S^z_{\bf q}(T)$ as shown in Fig.~\ref{check2}(c).
This means that the OFTLM is not only a highly accurate method but also a user-friendly method because one can choose $M$ without checking the convergence of eigenvalues in each Lanczos sampling.
 
\section{results}
\label{Sec4}

\subsection{Conditions of numerical calculation}
In the present study, we calculate $C(T)$, $\mathcal{S}(T)$, and $S^z_{\bf q}(T)$ using the RFTLM for $N=36$ and the OFTLM for $N=24$ and $N=30$.
The $N=24$, $N=30$, and $N=36$ clusters are shown in Fig.~\ref{lattice} for the KL and TL.
Finite-size effects can be reduced by using large-size and highly symmetric clusters such as $N=36$.
We emphasize that the improved FTLMs with high accuracy make finite-size effects at low temperatures very clear.

To calculate the excited states required for using the improved FTLMs, we use the restarted Lanczos method with the root-shifting method.
Table~\ref{para} shows detailed conditions for improved FTLM calculations.

For large clusters such as $N=36$, it is time-consuming to prepare several eigenvectors with $N_E>1$ or $N_V>1$. Furthermore, one has to be careful regarding the appearance of the ghost eigenvalues in such a huge calculation. To avoid these difficulties, we decide to use the RFTLM with $N_E=1$, where we need to calculate the ground state only. The accuracy of the $N_E=1$ result will be confirmed in the next section.

\begin{table}[tb]
\caption{Conditions for the improved FTLMs in our calculations.}
\begin{tabular}{|c|c|c|c|c|} \hline 
    $N$  &  Method & $R$ & $M$ & $N_E$ or $N_V$ \\ \hline 
 \  \  24 \  \  & \ \ OFTLM \ \ & 100 & 100-160 & 10  \\ \hline
    30 & OFTLM & 100 & 100-300 & 4   \\ \hline
    36 & RFTLM & \ 50-75 \ & \ 150-400 \ & 1  \\ \hline
  \end{tabular}
\label{para}
\end{table}

\subsection{Kagome lattice}

\begin{figure}[tb]
 \begin{center}
\includegraphics[width=86mm]{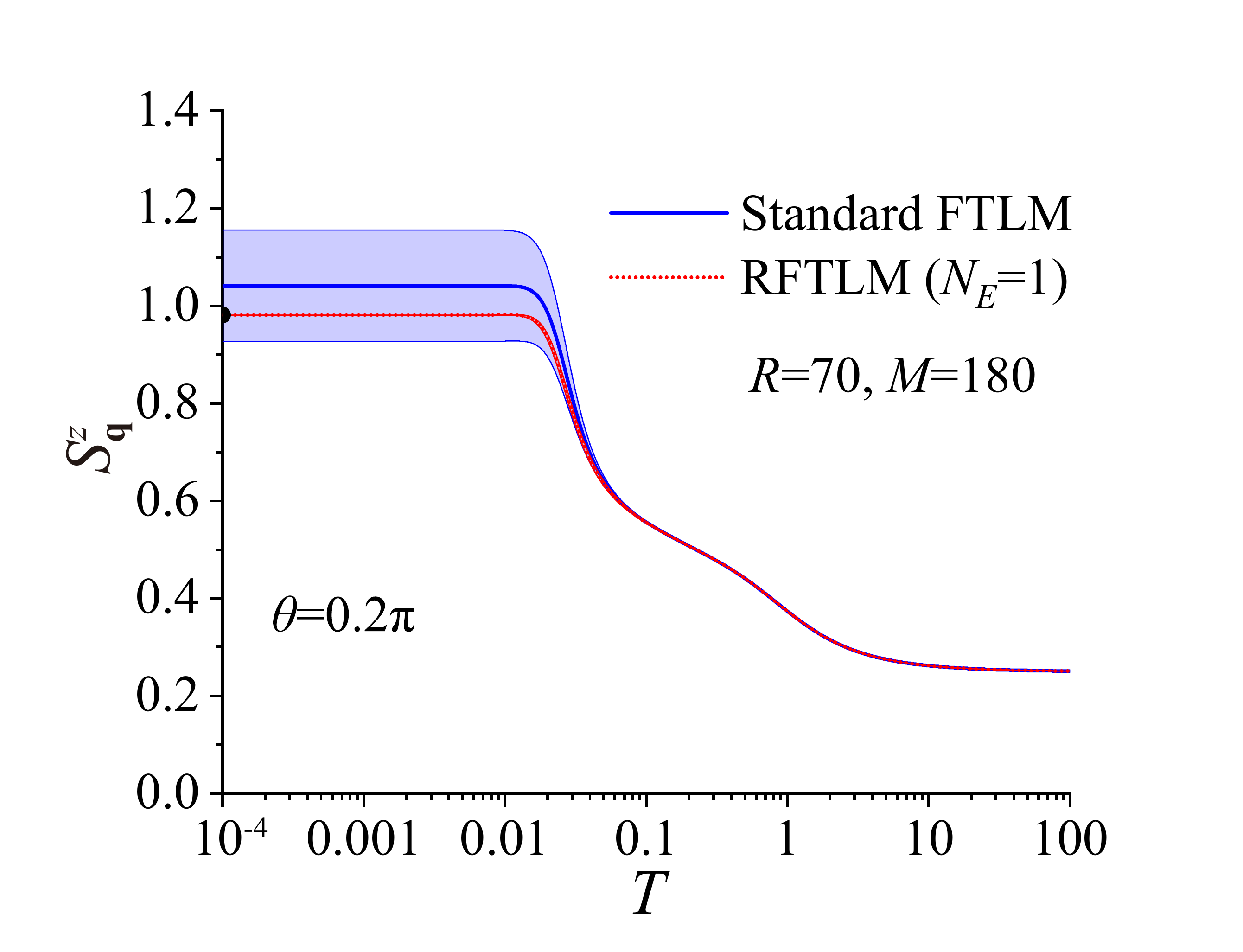}
   \caption{ Comparison between the standard FTLM and RFTLM for the accuracy of $S^z_{\bf q}(T)$ on the $N=36$ kagome system at $\theta=0.2\pi$. 
The blue shaded region indicates the standard errors of the FTLMs using the jackknife technique.
A black dot denotes the exact value at $T=0$.}
  \label{check3}
\end{center}
\end{figure}

We first discuss the efficiency of the RFTLM for the $N=36$ kagome system at $\theta=0.2\pi$.
Figure~\ref{check3} shows $S^z_{\bf q}(T)$ at ${\bf q}=(2\pi, 2\pi/\sqrt{3})$ using the standard FTLM and RFTLM. 
In the standard FTLM, there is large error at low temperatures and an average value at $T=0$ deviates from the exact one. 
On the other hand, in the RFTLM, the error bars become less than the width of the line for all temperatures and an average value converges to the exact one at $T=0$.
This clearly demonstrates that our improved FTLMs work well even for the $N=36$ system.
We emphasize that the error of the FTLMs becomes almost less than the line width in all the results shown below.

\begin{figure}[tb]
 \begin{center}
	\includegraphics[width=80mm]{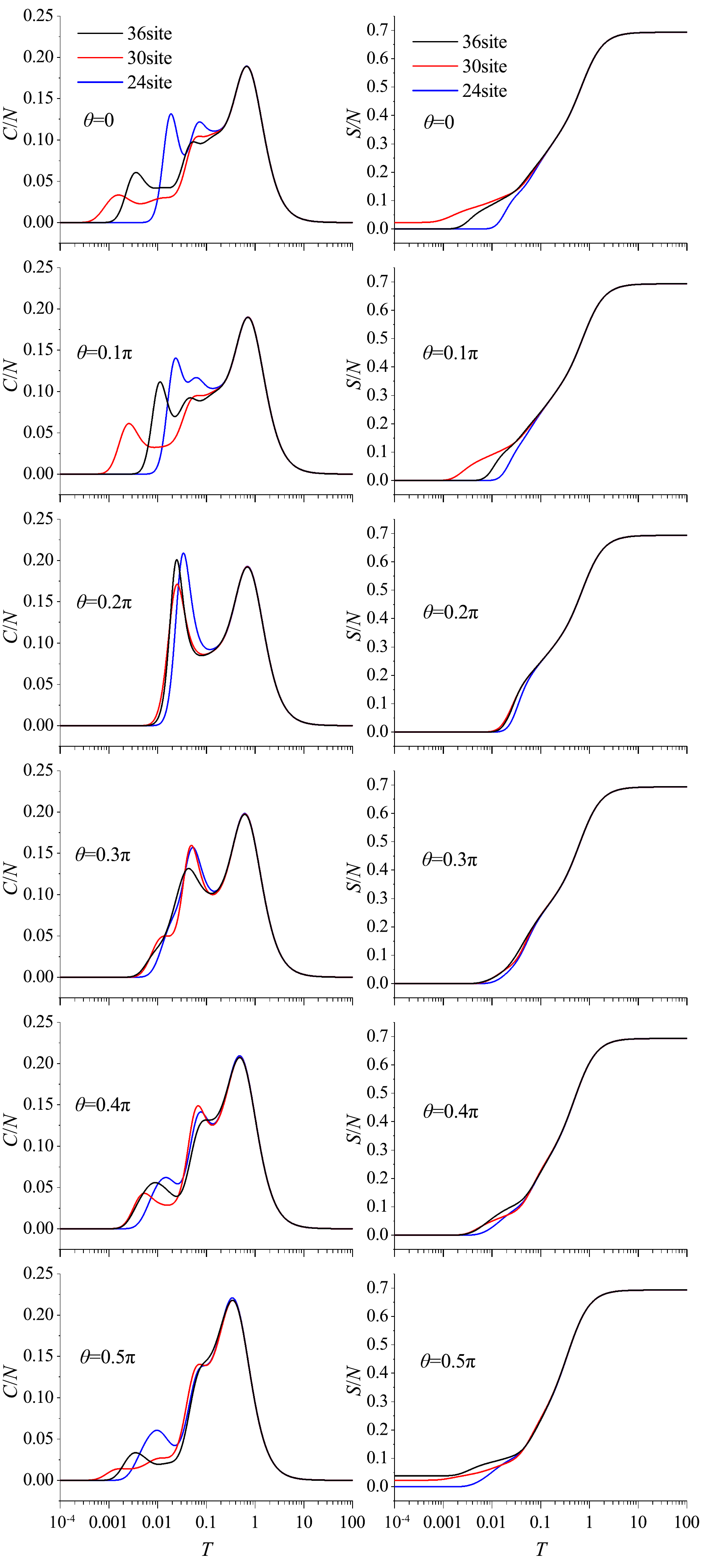}
   \caption{Temperature dependence of the specific heat $C$ (left panels) and entropy $\mathcal{S}$ (right panels) per site for the kagome system, obtained by using the RFTLM for $N=36$ and OFTLM for $N=24$ and $N=30$.
 Note that standard errors of the FTLMs are almost less than the line width.}
  \label{kagocs}
\end{center}
\end{figure}

\begin{figure}[tb]
 \begin{center}
	\includegraphics[width=86mm]{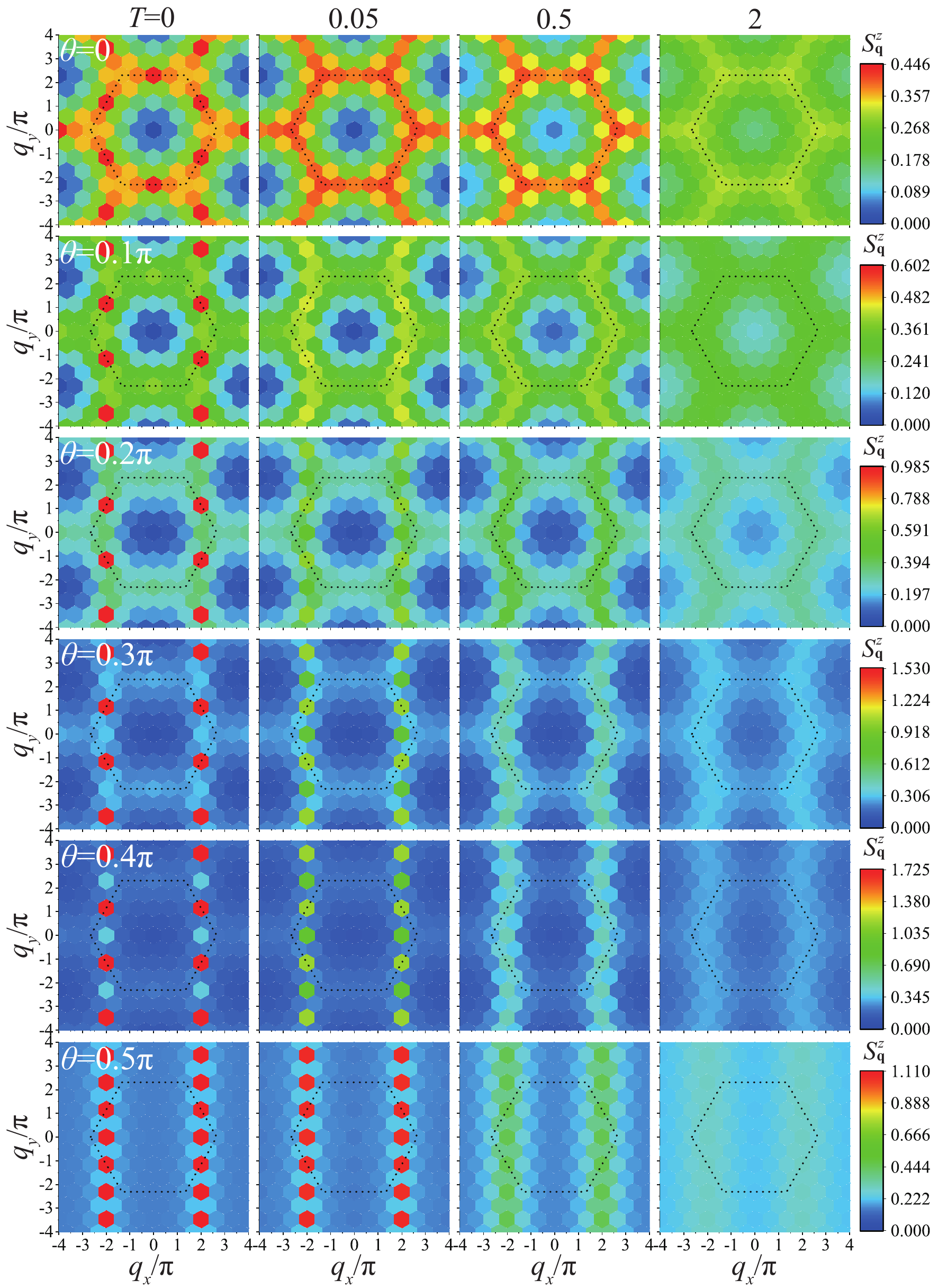}
   \caption{Color plots of the finite-temperature static spin structure factor $S^z_{\bf q}(T)$ for the $N=36$ kagome system, obtained by using the RFTLM. The black dotted hexagons denote the extended first Brillouin zone. The unit of length is the length of a side in the unit cell.}
  \label{kagosq}
\end{center}
\end{figure}

Figure \ref{kagocs} shows the calculated results of $C(T)$ (left panels) and $\mathcal{S}(T)$ (right panels) for $0 \leq \theta \leq 0.5\pi$ at $N=24$, 30, and 36.
 $C(T)$ exhibits the multiple-peak structures in all $\theta$ and $N$. 
For $T>0.2$ and all $\theta$, $C(T)$ is almost size independent.
Therefore, it is expected that a highest-temperature peak at $T\sim0.5$ shown in Fig.~\ref{kagocs} hardly changes even in the thermodynamic limit.

At $\theta=0$ and $\theta=0.1\pi$, we obtain two or three peaks for $T<0.2$ in all sizes. This is consistent with the previous studies for $\theta=0$~\cite{kft1,kft2,kft3,kft4,kft6}.
These low-temperature peaks are strongly size-dependent, and thus $C(T)$ in the thermodynamic limit is still unresolved.

At $\theta=0.2\pi$, $C(T)$ exhibits a clear double peak, which has hardly any difference between $N=30$ and $N=36$.
Therefore, the existence of this double peak is strongly expected even in the thermodynamic limit at $\theta=0.2\pi$.
In addition, the entropy shows a tendency toward a plateau around $\mathcal{S}(T)/N\sim0.3\sim\ln(2)/2$ shown in Fig.~\ref{kagocs}.
The plateau with $\mathcal{S}(T)/N=ln(2)/2$ has been obtained in the Kitaev model on a honeycomb lattice~\cite{hkdp,hkas1} and in RuCl$_3$ known as the Kitaev-like model compound~\cite{RC}. 
However, the origin of the plateau is different, which will be discussed in Sec.~\ref{Sec5}.

At $\theta=0.5\pi$ (Kitaev limit), $\mathcal{S}(T)$ for $N=30$ and $N=36$ becomes finite at the lowest temperature ($T=0.0001$), being consistent with twofold (fourfold) degeneracy in the ground state for $N=30$ (36). 
This degeneracy is partially consistent with a previous result using the cluster mean-field method~\cite{kkh}, predicting $2^{3L}$-fold degeneracy in the thermodynamic limit ($L\rightarrow\infty$), where $L$ is the linear system size giving the total lattice sites $N=3\times L^2$.

To explore the origin of the multiple-peak structure in $C(T)$, we calculate $S^z_{\bf q}(T)$ for $N=36$ by using the RFTLM, and the results are shown in Fig.~\ref{kagosq}. 
When $S^z_{\bf q}(T)$ has the largest intensity at the corner (the edge center) of the extended first Brillouin zone, a $\sqrt{3} \times \sqrt{3}$ state (a $\bf q=0$ state) appears with short-range order (SRO).
At $\theta=0$ (Heisenberg limit), we obtain a crossover from the paramagnetic state to the $\sqrt{3} \times \sqrt{3}$ SRO, and to the $\bf q=0$ SRO state, from high to low temperatures. This is the same result obtained by Shimokawa and Kawamura by using the Hams-de Raedt method~\cite{kft4}.
At $\theta>0$ and $T=0.5$ where the high-temperature peak in $C(T)$ appears, we can see that $S^z_{\bf q}(T)$ has zigzag or linear distribution in intensity along the $q_y$ direction on $q_x/\pi=\pm2$.
This result indicates that the origin of the high-temperature peak is attributed to a crossover from the paramagnetic state to the SRO state with a zigzag or linear intensity distribution on the SSSF. 
At $0.1\pi\leq\theta<0.5\pi$ and $T\leq0.05$, $S^z_{\bf q}(T)$ has the strongest intensity at the edge centers. 
Therefore, we expect that one of the lower-temperature peaks in $C(T)$ is a signature of the $\bf q=0$, $120^\circ$ order. 
At $\theta=0.5\pi$ (the Kitaev limit), the intensity distribution of $S^z_{\bf q}(T)$ has a perfect linear structure. 
This structure has been obtained in the classical spin system using the Monte Carlo method~\cite{kkh}.
This comes from the fact that there is a $120^\circ$ structure in every triangle of the KL but no clear correlation between neighboring triangles. The same can be expected for the quantum system.

Therefore, we can conclude that the order by disorder phenomenon does not occur even in the existence of both the quantum and thermal fluctuations.

\subsection{triangular lattice}
\begin{figure}[tb]
 \begin{center}
	\includegraphics[width=86mm]{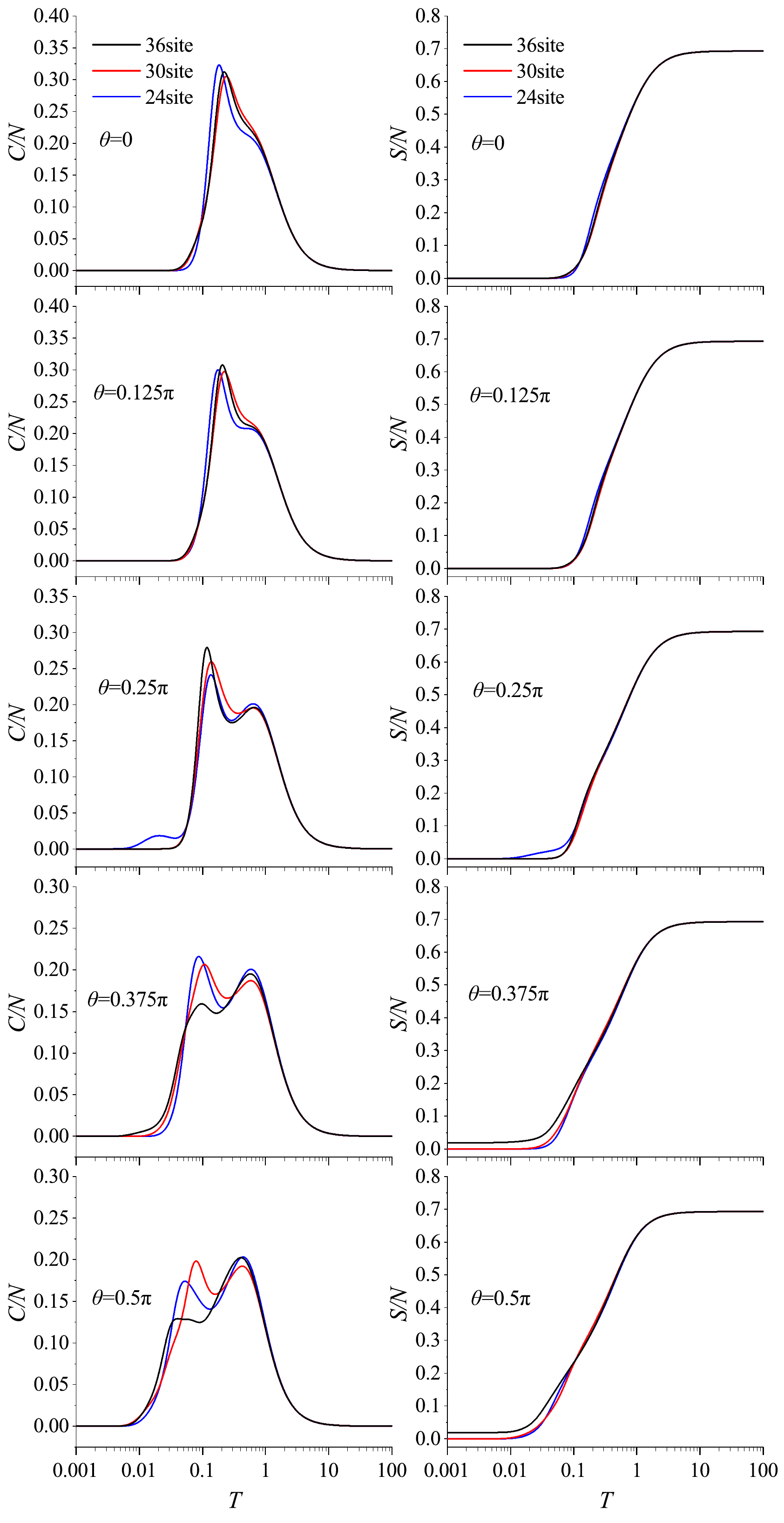}
   \caption{Temperature dependence of the specific heat $C$ (left panels) and entropy $\mathcal{S}$ (right panels) per site for the triangular system, obtained using the RFTLM for $N=36$ and OFTLM for $N=24$ and $N=30$.
 Note that standard errors of the FTLMs are almost less than the linewidth.}
  \label{trics}
\end{center}
\end{figure}

The classical ground states in the TL are predicted to be the $Z_2$ vortex crystal state and the nematic state in $0\leq\theta\leq0.5\pi$~\cite{tkh2,tkh3,tkh4,tkh5,tkh6}.
We perform finite-temperature calculations for the quantum triangular system.
In a recent study, it has been predicted that $C(T)$ at $\theta=0$ (Heisenberg limit) has two anomalies at $T\sim0.2$ and $T\sim0.55$~\cite{tft3}.
In our calculated $C(T)$ at $\theta=0$, a clear peak is obtained at $T\sim0.2$, and a shoulder-like anomaly is obtained at $T\sim0.6$, shown in Fig.~\ref{trics}.
A good agreement with the previous work corroborates the validity of our method.
In addition, 
we obtain a gradual change from the shoulder-like anomaly to a peak as $\theta$ is increased keeping the temperature unchanged.
On the other hand, at $T\sim0.2$ and $\theta=0.5\pi$ for $N=36$, the low-temperature peak structure is suppressed.
Since this peak exhibits a large-size effect, $C(T)$ at low temperatures in the thermodynamic limit still remains an unresolved problem.

The entropy of the triangular system is different from that of the kagome system, because there is no plateau-like anomaly in any $\theta$ and all $N$.
When $\theta\geq0.375\pi$ and $N=36$, the ground state has twofold degeneracy.
For this reason, the $\mathcal{S}(T)/N$ converges to a value of $\ln(2)/36$ at the lowest temperature $T=0.001$ as shown in Fig.~\ref{trics}.

\begin{figure}[tb]
 \begin{center}
	\includegraphics[width=86mm]{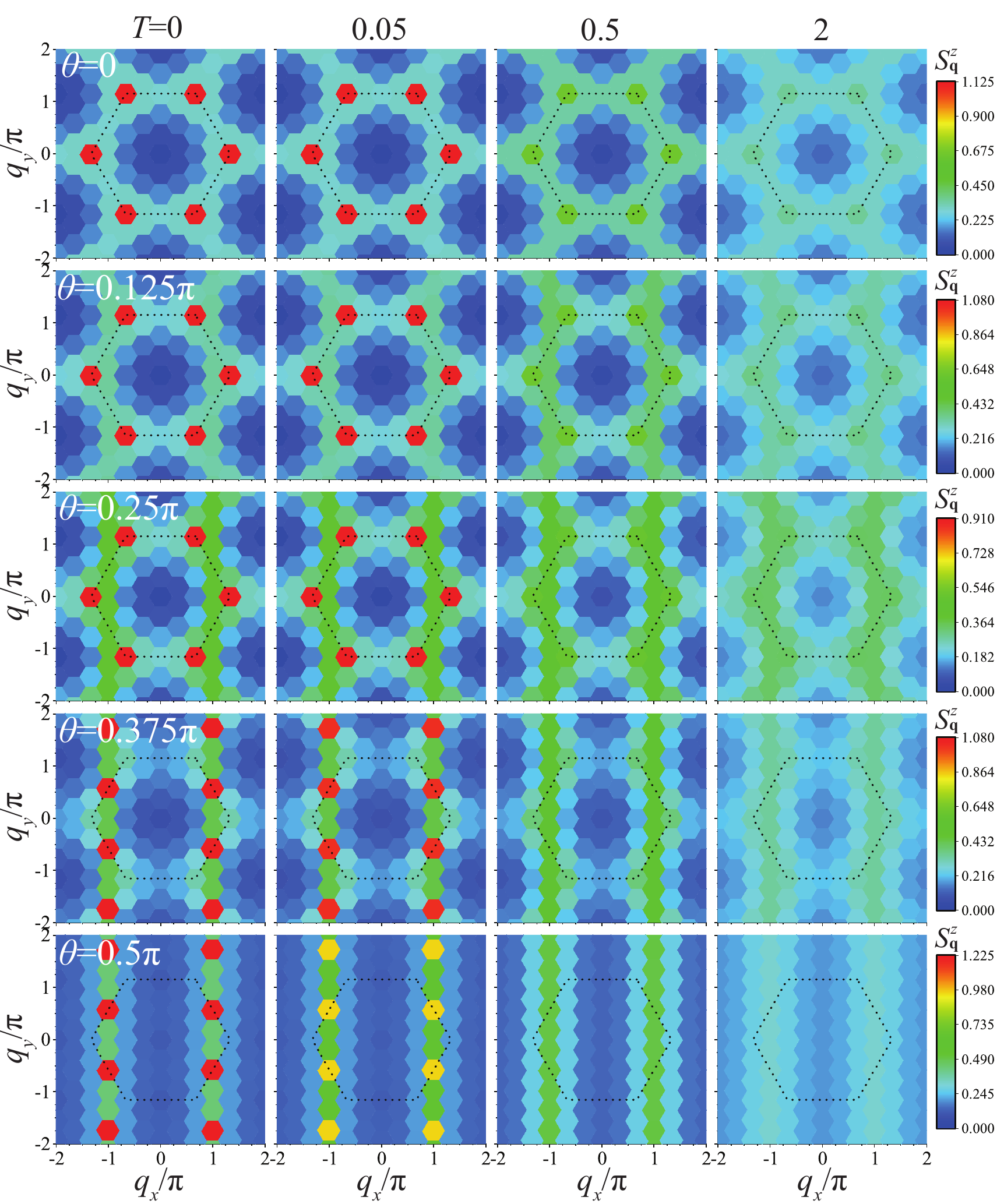}
   \caption{Color plots of the finite-temperature static spin structure factor $S^z_{\bf q}(T)$ for the $N=36$ triangular system, obtained by using the RFTLM. The black dotted hexagons denote the first Brillouin zone. The unit of length is the distance between nearest neighbors.}
  \label{trisq}
\end{center}
\end{figure}

\begin{figure}[tb]
 \begin{center}
	\includegraphics[width=86mm]{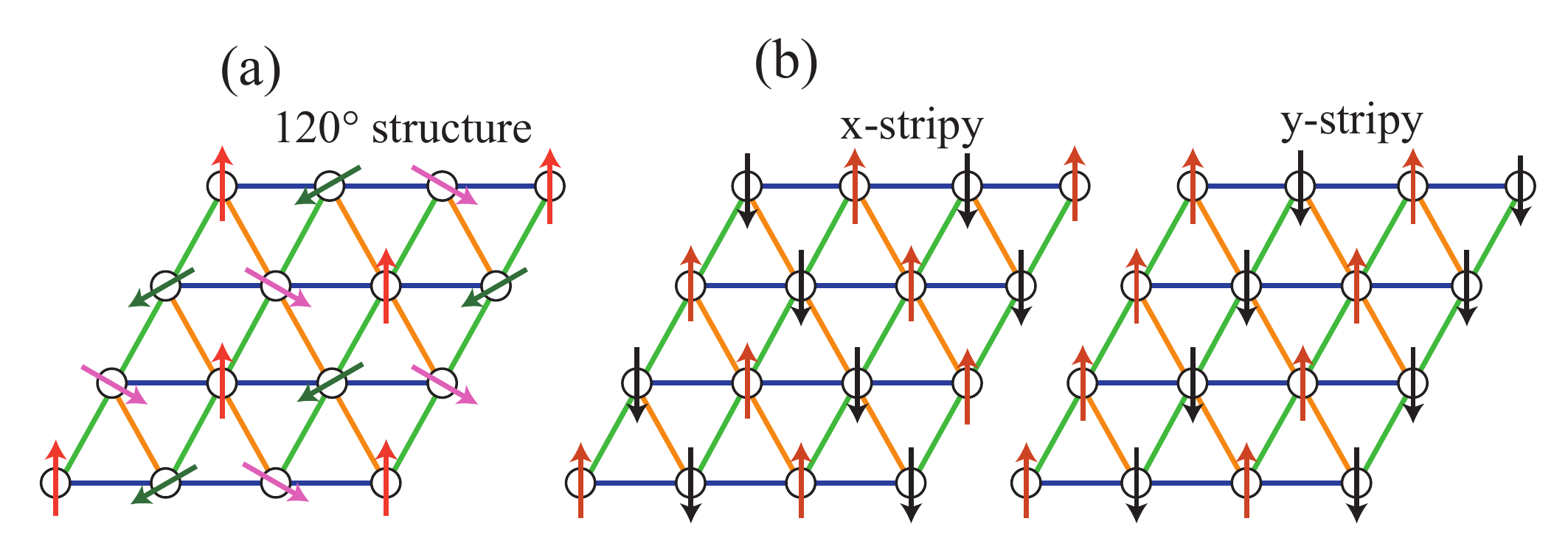}
   \caption{Schematic view of the ground states of the triangular system. (a) A 120$^\circ$ order state. (b) Stripy order states.}
  \label{tristr}
\end{center}
\end{figure}

We calculate $S^z_{\bf q}(T)$ of the triangular system for the $N=36$ cluster (Fig.~\ref{trisq}), which has a good rotational symmetry as shown in Fig.~\ref{lattice}(b). 
Similarly to the kagome system, for $\theta\geq0.25\pi$ the intensity distribution of $S^z_{\bf q}(T)$ exhibits a zigzag or linear structure along the $q_y$ axis on $q_x/\pi=\pm1$ at $T = 0.5$ where the high-temperature peak in $C(T)$ appears.
The linear-intensity distribution at $\theta=0.5\pi$ corresponds to a nematic state without long-range dipole order~\cite{tkh3}.
For this reason, the high-temperature peak in $C(T)$ is expected to be the signature of a crossover from the paramagnetic state to a nematic-like SRO state having a zigzag or linear structure as in the kagome system. 

Next we focus on $S^z_{\bf q}(T)$ at $T=0$.
At  $\theta=0$, $S^z_{\bf q}(0)$ has maximum intensity at the corners of the Brillouin zone, which corresponds to the 120$^\circ$ order as shown in Fig.~\ref{tristr}(a). 
The existence of the 120$^\circ$ order is consistent with other studies.
At $\theta=0.5\pi$, $S^z_{\bf q}(0)$ has maximum intensity at ${\bf q}=( \pi, \pi/\sqrt{3})$ and ${\bf q}=( \pi, -\pi/\sqrt{3})$, meaning the $x$-stripy order and $y$-stripy order, respectively, as shown in Fig.~\ref{tristr}(b).  
In the classical system, the ground state has a linear intensity distribution in the SSSF, which is nematic~\cite{tkh3}.
Therefore, we believe that the order by disorder phenomenon occurs in the $S=1/2$ TL Kitaev model due to the quantum fluctuation.
This order has been predicted in the analysis by the linked-cluster expansion and spin-wave theory~\cite{tko}.

For $0<\theta<0.5\pi$, we cannot find evidence of the $Z_2$ vortex crystal state that has a multiple-$q$ structure in the SSSF, probably because of the limited system size.
Nevertheless, we believe that there is long-range order (LRO) or SRO related to the $Z_2$ vortex crystal state at low temperature in the thermodynamic limit, as in the classical system.

\section{Discussion}
\label{Sec5}
We compare the results of the kagome and triangular KH model with the honeycomb Kitaev model. 
In the honeycomb Kitaev model, it has been elucidated that $C(T)$ has a double-peak structure.
In the kagome and triangular KH model, we have found the multiple-peak structures in this work.
However, the origins of the double-peak and multiple-peak structures different.
In the honeycomb Kitaev model, the double peak is caused by the itinerant Majorana fermions and $Z_2$ fluxes freezing at different temperatures~\cite{hkdp}.
In the kagome system at $0<\theta<0.5\pi$, 
the high-temperature peak is a consequence of a crossover from the paramagnetic state to a SRO state whose SSSF has a zigzag or linear intensity distribution, and one of the low-temperature peaks has been expected to be a signature of the $\bf q=0$, $120^\circ$ order. 
At $\theta=0.5\pi$, there is only crossover from the paramagnetic state to a SRO state whose SSSF has a linear intensity distribution. This linear intensity distribution comes from the fact that there is a $120^\circ$ structure in every triangle of the KL but no clear correlation between neighboring triangles.
Therefore, we believe that at $0<\theta<0.5\pi$ there are two or more peaks in $C(T)$ in the thermodynamic limit, whereas at $\theta=0.5\pi$ there is only one peak. 
 
In the triangular system, 
the high-temperature peak at $\theta>0.25\pi$ has the same origin as the kagome system.
At $0<\theta<0.5\pi$, we can expect that there is a low-temperature peak in $C(T)$ because of the LRO or SRO related to the $Z_2$ vortex crystal state in the thermodynamic limit.
At $\theta=0.5\pi$, the high-temperature peak is a consequence of the crossover from the paramagnetic state to a nematic-like SRO state, while the low-temperature peak is a signature of the stripe LRO.

Because of the emergence of LRO and/or SRO due to the Heisenberg term, a peak on the low-temperature side of $C(T)$ develops with decreasing $\theta$ in both the KL and TL. Therefore, we can say that there is a competitive effect between the Heisenberg and Kitaev terms with respect to the intensity of the low-temperature peak in $C(T)$.

The kagome and triangular systems have a significant difference at $\theta=0.5\pi$ (Kitaev limit). 
In the triangular system, order by disorder due to the quantum fluctuations occurs in common with many frustrated quantum spin systems, and the ground state becomes the stripe order.
On the other hand, it does not occur in the kagome system. 

We have developed new improved FTLMs: these are the RFTLM and OFTLM.
These FTLMs improve the accuracy for all physical quantities at low temperatures compared to the standard FTLM.

\section{Summary}
\label{Sec6}
Inspired by the remarkable development of the quantum Kitaev-Heisenberg models in recent years,
we investigated the finite-temperature properties of the $S=1/2$ KH models on the kagome lattice and triangular lattice by means of improved finite-temperature Lanczos methods.
We obtained the multiple peaks in the specific heat in both lattice models.
The origin of the high-temperature peak of the specific heat is attributed to a crossover from the paramagnetic state to the SRO state with a zigzag or linear structure on the SSSF.
We believe that the origin of the low-temperature peak is the $\bf q=0$, $120^\circ$ order in the KL and the $Z_2$ vortex state in the TL, caused by the Heisenberg term.

We also reveal that at $\theta=0.5\pi$ (Kitaev limit) in the triangular system, the ``order-by-disorder'' phenomenon due to the quantum fluctuations occurs, and the ground state exhibits the stripe order.
On the other hand, in the kagome system it does not occur even in the presence of both the temperature and quantum fluctuations.
We believe this effect is peculiar to the Kitaev model on the kagome lattice.

We have succeeded in improving the finite-temperature Lanczos method.
For larger systems, we can expect further improvements, especially faster calculations, using a technique for decomposing full Hilbert space with several symmetries such as in the case of SPINPACK~\cite{sp}.
The next target for finite-temperature calculations will be lattices with 48 sites, which remains a future work.

\begin{acknowledgments}
This work was supported by MEXT, Japan, as a social and scientific priority issue (creation of new functional devices and high-performance materials to support next-generation industries) to be tackled by using a post-K computer. The numerical calculation was carried out at the facilities of the Supercomputer Center, the Institute for Solid State Physics, the University of Tokyo and at the Yukawa Institute Computer Facility, Kyoto University. 
\end{acknowledgments}

\end{document}